\documentclass[pra,reprint,amsmath,amssymb,aps,onecolumn]{revtex4-2}

\usepackage{graphicx}
\usepackage{dcolumn}
\usepackage{bm}
\usepackage[dvipsnames]{xcolor}
\usepackage{datetime}
\usepackage{color}
\def\be{\begin{eqnarray}}
\def\ee{\end{eqnarray}}

\def\({\left(}
\def\){\right)}

\def\k10{k_{1\rightarrow 0}}
 
\begin{document}


\title{Velocity Map Imaging with No Spherical Aberrations}%

\author{Yehuda Ben-Shabo$^1$, Adeliya Kurbanov$^1$, Claus Dieter Schr{\"o}ter$^2$, Robert Moshammer$^2$, Holger Kreckel$^2$, Yoni Toker$^1$}

\affiliation{$^1$Department of Physics and Institute of Nanotechnology and Advanced Materials, Bar-Ilan University, Ramat-Gan 5290002, Israel\\
$^2$Max-Planck-Institut f\"ur Kernphysik, Saupfercheckweg 1, 69117 Heidelberg, Germany\\}

\begin{abstract}
Velocity map imaging (VMI) is a powerful technique that allows to infer the kinetic energy of
ions or electrons that are produced from a large volume in space with good resolution. The size
of the acceptance volume is determined by the spherical aberrations of the ion optical system.
Here we present an analytical derivation for velocity map imaging with no spherical aberrations.
We will discuss a particular example for the implementation of the technique that allows using the
reaction microscope recently installed in the Cryogenic storage ring (CSR) in a VMI mode. SIMION
simulations confirm that a beam of electrons produced almost over the entire volume of the source
region, with width of 8 cm, can be focused to a spot of 0.1 mm on the detector. The use of the same
formalism for position imaging, as well as an option of position imaging in one axis and velocity
map imaging in a different axis, are also discussed. 
\end{abstract}

\maketitle

\section{Introduction}
The goal of photo-electron spectroscopy (PES) is to measure the velocity components of photo-emitted electrons in order to determine the energy of the atomic/molecular orbitals \cite{Weichman2018, Mabbs2009}, their symmetry \cite{Cooper1968,Reid2003} and excited state dynamics \cite{Stolow2004}. In imaging PES this is achieved by accelerating the electrons towards a position-sensitive particle detector and measuring their 2-dimensional impact positions and, if possible, also their time-of-flight (TOF) \cite{Chichinin2009,Basnayake2022}. A similar approach is used in ion imaging to infer the kinetic energy release of ions produced in gas phase reactions \cite{Suits2018}. In these imaging measurements the three measured quantities (the time-of-flight and the impact positions on the detector) are the result of effectively 6 parameters: the three components of the particles initial velocity, and the three components of the particles position of origin (which in this case are not of interest). There are two main strategies for avoiding the broadening of the measured spectra due to the uncertainty in the particles source point. The first is to create the electrons/ions from a small region in space, for example by tightly focusing a photo-ionizing laser. The second approach, known as velocity map imaging (VMI, also known as 'position focusing'), which was first described by Eppink and Parker in 1997 \cite{Eppink1997}, is to use electrostatic focusing to ensure that the impact position on the detector does not depend on the particles original position, only on the particles initial velocity. An intuitive understanding of this approach builds on the equivalence between optics and ion optics. In optics, when the detector is positioned in the focal plane of the lens (also known as the Fourier plane) parallel rays of light are focused to a point, i.e., the impact position does not depend on the size or extension of the light source, only on its directionality. Correspondingly, in ion optics, VMI is achieved using an electrostatic lens whose voltage is set so that its focal plane lies on the detector. Notably, the technical constraints in ion optics are different. Here, the main constraint is that the electric potential, $V(x,y,z)$ must satisfy the Laplace equation:
\begin{equation} \label{E:Laplace}    
\nabla^2V=\frac{\partial^2V}{\partial x^2}+\frac{\partial^2V}{\partial y^2}+\frac{\partial^2V}{\partial z^2}=0.
\end{equation}
In practice, the design of ion optical devices usually adds additional constraints. For example, one would like to use only a small number of electrodes and power supplies. Most VMI designs use a large field free region, or a region with a constant electric field\cite{Horke2012}. The original work by Eppink and Parker \cite{Eppink1997} employed only three electrodes (a repeller, an extractor and a ground electrode), however it soon became clear that one can reduce spherical aberrations by adding more electrodes \cite{Suits2018}. A comprehensive review of the different implementations of VMI is beyond the scope of this paper, we will only note that configurations with a large number of electrodes spanning the entire volume from the interaction region to the detector (known as 'thick lens' designs) are increasingly being used \cite{Lin2003,Horke2012, Kling2014, Marchetti2015, Ding2021,Wu2023, Davino2023}. However, as the complexity increases, finding the optimal design and voltages becomes more challenging requiring extensive iterative simulations \cite{Marchetti2015}. According to a recent work \cite{Wu2023}, the current state of the art enables focusing a beam with a width of a few mm in the on source region to a $\simeq 0.1~$mm spot on the detector, resulting in a 'focusing factor' (also known as a 'blurring factor') of $\simeq 0.05$ \cite{Wu2023}. 

In the following we will analyze charged particle trajectories in a quadratic electrostatic potential and demonstrate how it can be used for VMI with no spherical aberrations. We discuss a practical implementation of this methodology as a method of implementing photoelectron spectroscopy in the Cryogenic Storage Ring (CSR), located at the Max Planck Institute for Nuclear Physics (MPIK), Heidelberg, Germany. The CSR is a purely electrostatic ion storage ring with a circumference of $35~$m, which can be cooled down to a few degrees kelvin. The primary goal of the CSR is to study laboratory astrophysics as for the storage of ions in conditions of extremely low pressure and cryogenic temperature similar to the conditions in the interstellar medium, and to study astrophysically relevant reactions including the interaction of cold molecular ions with photons, electrons and neutral atoms. \cite{CSR_2016_RSI} 

Recently, a reaction microscope (CSR-ReMi) was installed in one of the straight sections of the CSR. Reaction microscopes (also known as COLTRIMS) use a combination of electric and magnetic fields and two opposing particle detectors to image, in coincidence, both electrons and charged fragments resulting from chemical reactions. Consequently, the kinetic energy of the electrons, as well as the mass and kinetic energy of the charged fragments is determined, resulting in a full kinematic description of the reaction \cite{Dorner2000_Rev,Ullrich2003Rev,HSB2021_Rev}. We present below SIMION simulations \cite{SIMION_Dahl2000} of how the CSR-ReMi can be operated in a VMI mode. We show that using the analytically derived voltages, with no further optimization, one can focus electrons from most of the inner diameter of the electrodes down to a spot size smaller than the spatial resolution of the detector, resulting in a focusing factor that is $40$ times smaller than the current state of the art\cite{Wu2023}. In addition, we discuss how the same methodology can be used for position imaging and in a mixed mode in which there is position imaging along one axis and velocity map imaging along the other axis.

\section{VMI with a quadratic electric potential}
Let us denote the spectrometers main axis (i.e. the direction from the interaction point to the middle of the detector) as the $\widehat{z}$ direction, and the $\widehat{x},\widehat{y}$ plane as the 'transverse plane'. Examine the following quadratic solution of the Laplace equation (Eq. \ref{E:Laplace}):
\begin{equation}\label{E:V}
V(\rho,z)=U_\rho(\rho^2-2z^2)-U_z z 
\end{equation}
where $V$ is the electrostatic potential, $\rho=\sqrt{x^2+y^2}$, and $U_\rho,~U_z$ are constants. 

The equation of motion along the spectrometer's main axis for a particle with charge $q$ and mass $m$ is given by:
\begin{equation}\label{E:z EOM}
    \ddot{z}=2\omega_\rho^2z+\frac{q}{m}U_z
\end{equation}
with the solution:
\begin{equation} \label{E:z solution}
    z(t)=\left(z_0+\frac{U_z}{4U_\rho}\right) cosh\left(\sqrt{2}\omega_\rho t \right)+\frac{v_z^0}{\sqrt{2}\omega_\rho} sinh\left(\sqrt{2}\omega_\rho t\right)-\frac{U_z}{4U_\rho}.
\end{equation}
Here $z_0,~v_z^0$ are the initial position and velocity along the $\widehat{z}$ axis, respectively. The time-of-flight, $t_f$, is set by $z(t_f)=L$, where $L$ is the distance to the detector. The equation of motion in the transverse direction is:
\begin{equation} \label{E:rho EOM}
    \ddot{\rho}=-2\frac{q}{m}U_\rho \rho = -\omega_\rho^2\rho
\end{equation}
where:
\begin{equation}\label{E:omega rho}
\omega_\rho\equiv \sqrt{2\frac{q}{m}U_\rho}
\end{equation}
with the solution:
\begin{equation} \label{E:rho solution}
\rho(t)=\rho_0 cos\left(\omega_\rho t\right)+\frac{v_\rho^0}{\omega_\rho} sin\left(\omega_\rho t\right).
\end{equation}
Here $\rho_0,~v_\rho^0$ are the initial position and velocity in the transverse direction, respectively. We denote the impact position on the detector by $\rho_f=\rho(t_f)$. Ideal VMI conditions correspond to no dependence of $\rho_f$ on $\rho_0$, which is achieved by setting $cos(\omega_\rho t_f)=0$, and consequently:
\begin{equation} \label{E:omega_rho VMI}
    \omega_\rho t_f=\pi/2+n\pi.
\end{equation}
where $n$ is an integer number. We will only treat the case of $n=0$. Inserting Eq. \ref{E:omega_rho VMI} into Eq. \ref{E:z solution}, we find that the condition for VMI is: 
\begin{equation}\label{E:VMI conditions}
    \frac{U_\rho}{U_z}=\frac{\beta_{vmi}}{L}
\end{equation}
with:
\begin{equation}
    \beta_{vmi}=\frac{1}{4}\left[cosh\left(\frac{\pi}{\sqrt{2}}\right)-1\right]=0.9161.
\end{equation}
We define the magnification, $M$, of the device as:
\begin{equation} \label{E:Magnification in VMI}
    M \equiv \frac{\rho_f}{v_\rho^0}
\end{equation}
In VMI conditions the magnification is given by:
\begin{equation}
    M=\frac{1}{\omega_\rho}=\sqrt{\frac{m}{2qU_\rho}}.
\end{equation}
The two parameters $U_\rho$ and $U_z$ allow us to set a desired magnification, $M$, while maintaining VMI conditions. Let $\epsilon_{max}[eV]=\frac{1}{2}m v_{max}^2$ be the maximal electron kinetic energy to be measured, then the desired magnification is given by:
\begin{equation}
    M=\frac{R_d}{v_{max}}=R_d\sqrt{\frac{m}{2\epsilon_{max}}},
\end{equation} 
where $R_d$ is the detector radius, leading to:
\begin{equation} \label{E:Us from emax}
U_\rho=\frac{\epsilon_{max}}{q R_d^2},~~U_z=\frac{L U_\rho}{\beta_{vmi}}=\frac{L\epsilon_{max}}{q\beta_{vmi}R_d^2}.
\end{equation}
Due to the cylindrical symmetry, to first order the position of impact, $\rho_f$ does not depend on $z_0$ and $v_z^0$. In the supplementary information the full expression for $t_f$ is derived (Eq. 11 of the supplementary information) and is used to derive the higher order corrections for $\rho_f$, showing that (Eq. 23 in the supplementary information):
\begin{equation}
    \rho_f=\frac{v_\rho^0}{\omega_\rho}+1.46 \frac{\rho_0z_0}{L}+\frac{0.9161}{\omega_\rho L}\rho_0 v_z^0+...
\end{equation}

\section{SIMION simulations of VMI imaging in the CSR-REMI}
\begin{figure}[ht] 
\centering
\includegraphics[width=.48\textwidth]{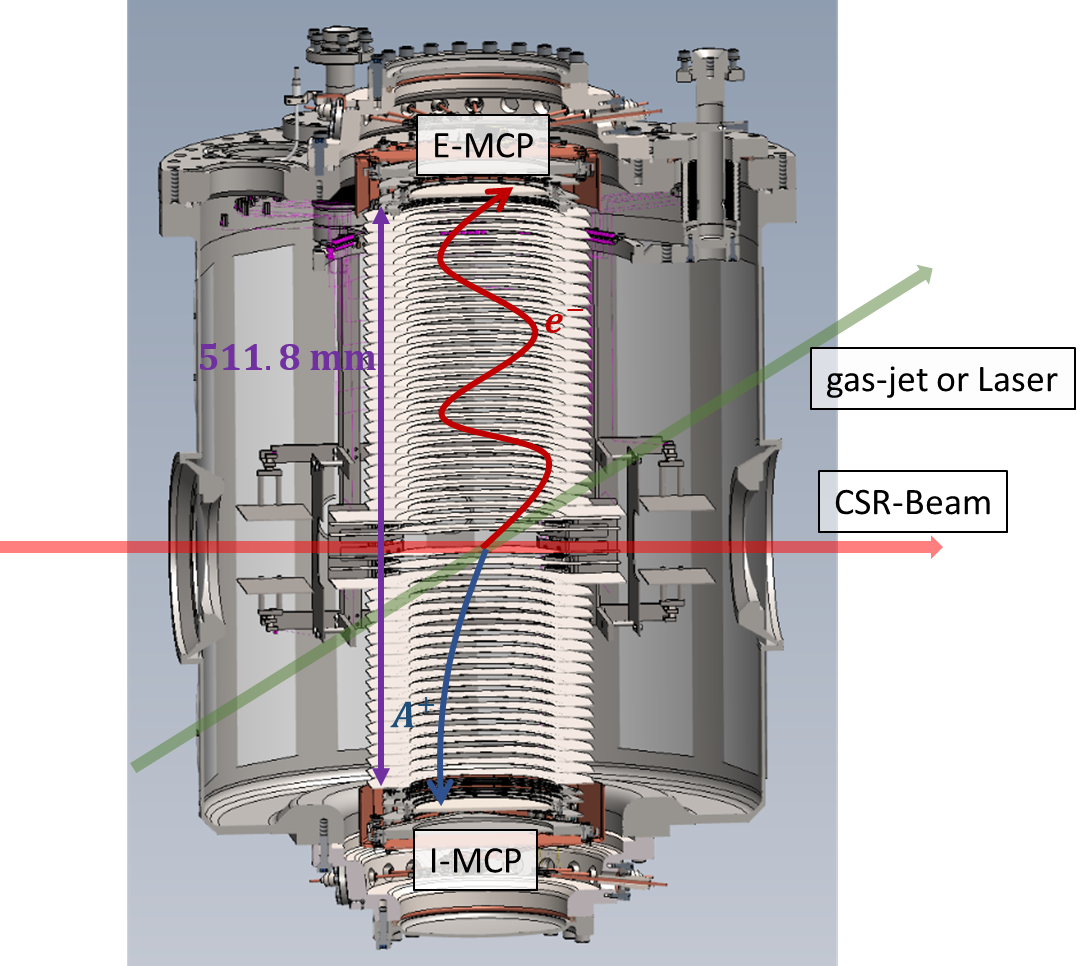}
\caption{A schematic illustration of the CSR-ReMi. In this illustration the stored ion beam passes through the instrument from left to right (along the direction indicated by the red arrow), and is crossed by either a neutral gas jet or by laser (along the direction indicated by the green arrow). Resulting electrons are accelerated upwards and imaged on the E-MCP detector, while resulting cationic fragments are accelerated in the opposite direction and imaged on the I-MCP detector.}
\label{F:CSR-ReMi}
\end{figure}
The CSR-ReMi, shown schematically in Fig. \ref{F:CSR-ReMi}, consists of 51 circular concentric electrodes, each with an inner diameter of $130~$mm, and microchannel plate (MCP) detectors on either side, one detector for the electrons (labeled E-MCP in Fig. \ref{F:CSR-ReMi}) and one for the cationic fragments (labeled I-MCP in Fig. \ref{F:CSR-ReMi}). The electrodes are connected through variable voltage dividers such that the voltage on each electrode can be set independently. The detectors have an active diameter of $120~$mm and are read out using delay-line anodes with a position resolution of $\simeq 0.16~$mm. The stored ion beam will pass through the center of the ReMi where it will be intersected with a neutral gas jet or with a laser beam. The primary use of the CSR-ReMi will be dissociative ionization experiments in which the neutral molecules from the gas jet will be ionized by collisions with the stored ions. Here, we discuss the possibility of using the CSR-ReMi for performing PES of the stored ions, and to study the kinetic energy of thermionically emitted electrons. The former case refers to the possibility of irradiating the stored ion beam by laser-light within the CSR-ReMi, and imaging the photo-electrons on the E-MCP detector. In this application, due to the low ion density and because high laser intensity is not required, one would prefer not to focus the laser to a small spot but rather have a large overlap between the laser and the stored ion beam. By thermionic emission we refer to electrons emitted through delayed electron emission, i.e., electrons emitted at very long times (ranging up to milliseconds) after photon absorption, or from ions that are in highly excited states caused by their production in the ion source. In this application VMI is necessary as the electrons are not emitted from a well-defined position. 

\begin{figure}[ht] 
\centering
\includegraphics[height=.6 \textwidth , width=.4 \textwidth]{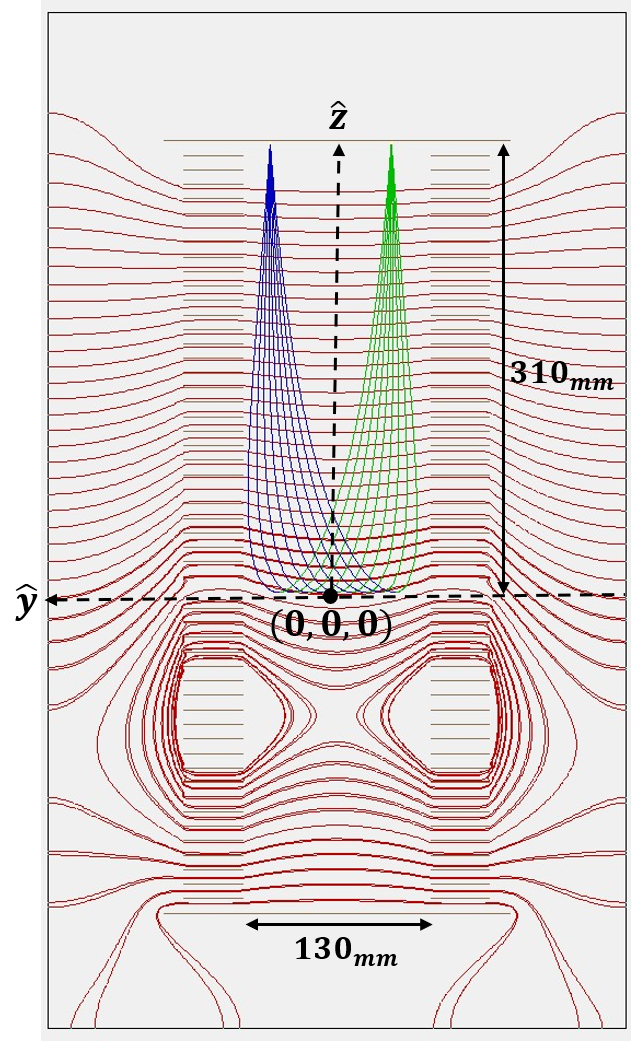}
\caption{A SIMION simulation of the CSR-ReMi, with equipotential lines marked in red. Also shown are electron trajectories. The blue trajectories correspond to a beam of electrons with width of $80~$mm on the $\widehat{y}$ direction and with a kinetic energy of $0.9~$eV in the $\widehat{y}$ direction. The green trajectories correspond to a kinetic energy of $0.6~$eV in the $-\widehat{y}$ direction. Both beams are focused to a spot size smaller than the position resolution of the detector ($92~\mu$m and $85~\mu$m in diameter, respectively).}
\label{F:Equipotential lines}
\end{figure}

Figure \ref{F:Equipotential lines} shows a SIMION simulation of the CSR-ReMi in VMI mode. In the simulations we set the voltages on the electrodes according to Eq. \ref{E:V} with no further optimization. Choosing a value of $\epsilon_{max}=1.8~$eV leads to the equipotential lines shown in red in Fig. \ref{F:Equipotential lines}. As can be seen, the potential has a saddle point located about $80~$mm below the interaction region. The calculated trajectories, for two monoenergetic electron beams are also shown. Here each electron beam has a width of $80~$mm along the $y$ axis, and both beams are focused to a spot size of less than $0.1~$mm (corresponding to a focusing factor of $0.0125$), i.e., smaller than the position resolution of the detector.

\begin{figure} 
    \centering
    \includegraphics[width=.4\textwidth]{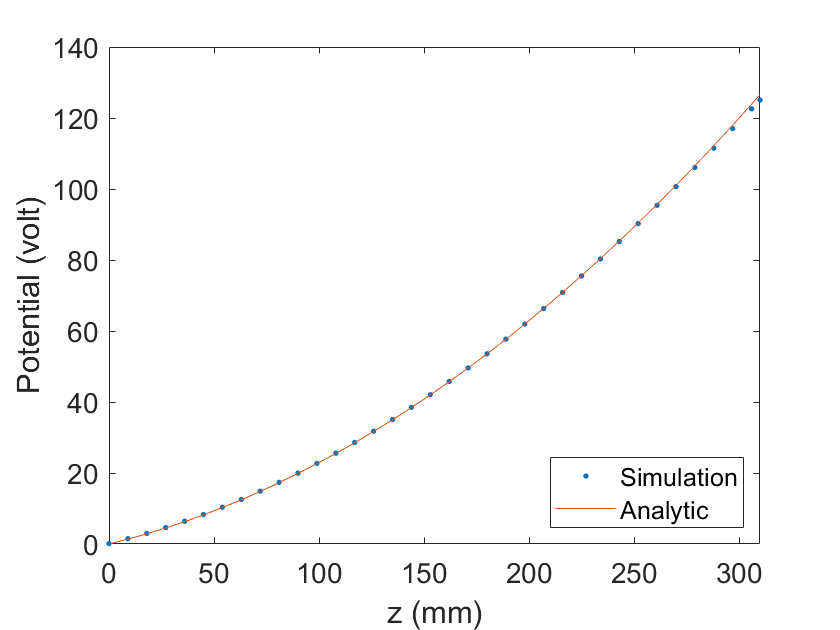}
    \caption{Comparison of the numerically calculated potential along the $\widehat{z}$ axis (for the same values of $U_\rho$ and $U_z$ used in Fig.\ref{F:Equipotential lines}), as a function of $z$, to the analytical formula (Eq. \ref{E:V}). Here $z=0$ corresponds to the center of the interaction zone. We observe small deviations (up to $1.2\%$) from the analytical potential only at small distances ($<15~$mm) from the detector.} 
    \label{F:Potential SIMION vs Theory}
\end{figure}

The fact that the final spot size on the detector has a width shows that there is a difference between the simulated and the analytical potentials, stemming from the finite size of the electrodes and from the fact that the detector is flat (and not in the shape of an equipotential line). However, the deviations from the analytical potential are small as can be seen in Fig. \ref{F:Potential SIMION vs Theory}, where a comparison of the calculated and analytical potential is shown. In the following we present a characterizations of the CSR-ReMi in VMI mode exhibiting a good match between the SIMION calculations and the analytical derivation.

\begin{figure} 
    \centering
    \includegraphics[width=.48\textwidth]{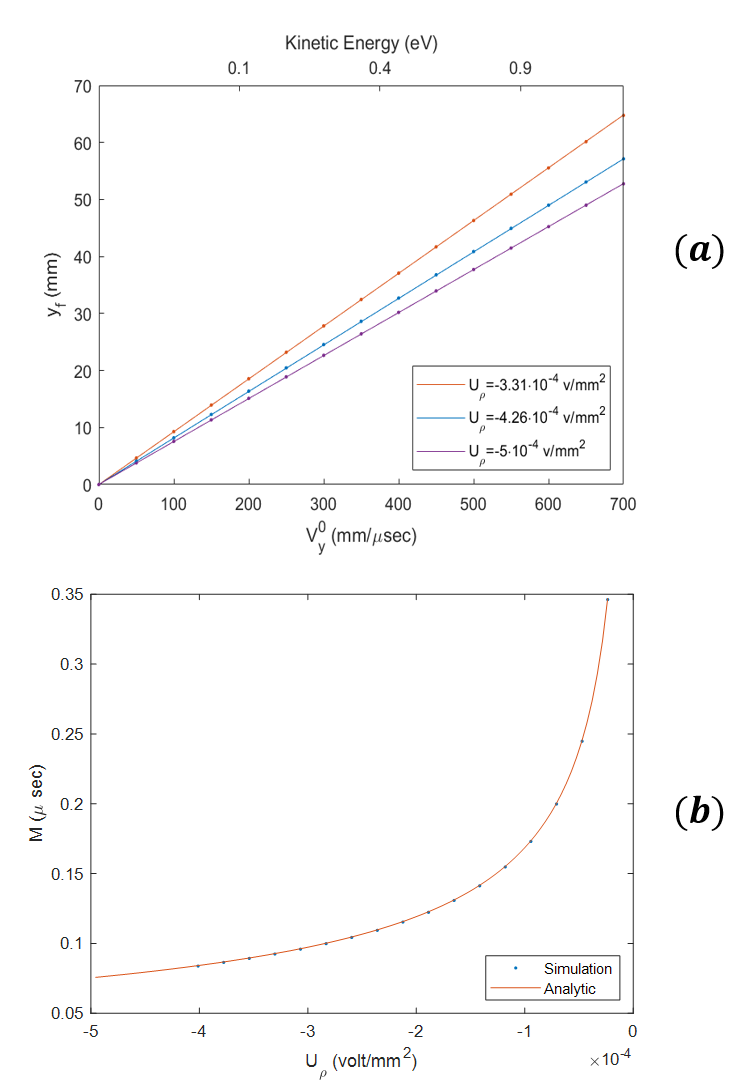}
    \caption{\textbf{(a)} The final position on the detector as a function of the initial velocity, for electrons that are starting from the origin with initial velocity on $y$ axis only, for different values of $U_\rho$. The continuous lines are given by the analytical derivation and the dots are given by the simulation. \textbf{(b)} The magnification as a function of $U_\rho$. The magnification is given by $\frac{1}{\omega_\rho}=\sqrt{\frac{m_e}{2qU_\rho}}$. For smaller absolute values of $U_\rho$, the magnification is larger.} 
    \label{F:rho vs vy}
\end{figure}

\begin{figure} 
    \centering
\includegraphics[width=.48\textwidth]{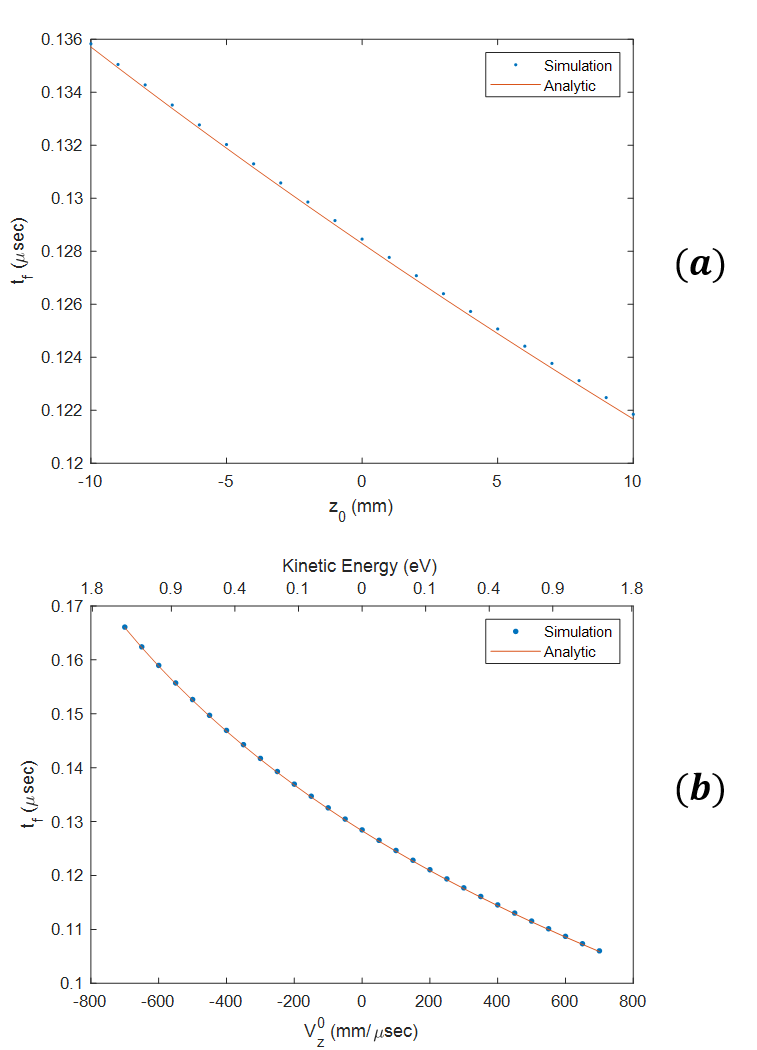}
    \caption{\textbf{(a)} Time-of-flight for a beam of electrons starting from different positions along the $\widehat{z}$ axis between $z_0=-10~$mm to $z_0=10~$mm, for $x_0=y_0=0$, and for values of $U_\rho=-4.26\cdot 10^{-4}~\mathrm{\frac{V}{mm^2}}$, $U_z=-0.1442~\mathrm{\frac{V}{mm}}$. \textbf{(b)} Time-of-flight for a beam of electrons starting from the origin with velocity on $z$ axis only, for the same values of $U_\rho$ and $U_z$ as for the left figure.} 
    \label{F:Time of Flight}
\end{figure}

\begin{figure} 
    \centering
\includegraphics[width=0.48\textwidth]{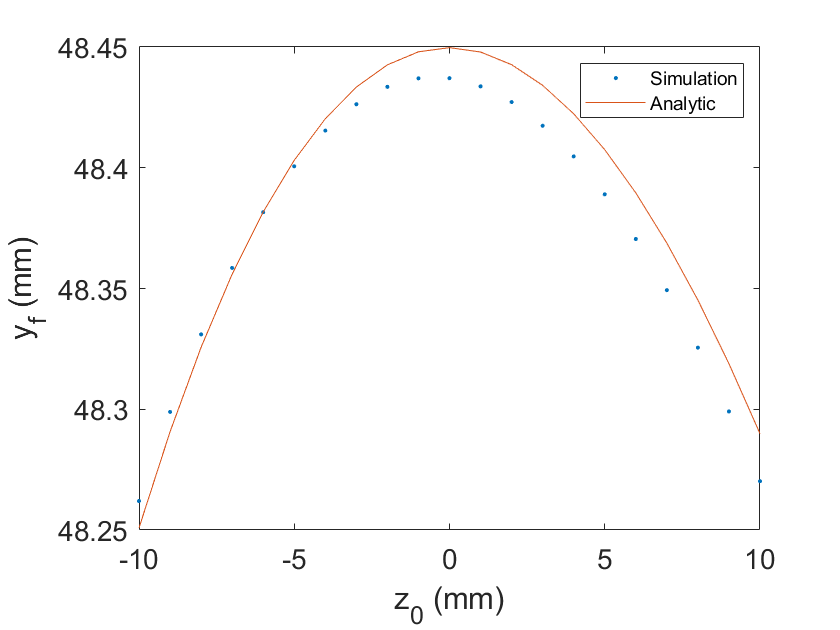}
    \caption{Impact position $y_f$ as a function of the source position $z_0$, for the same values of $U_\rho$ and $U_z$ as in Fig.\ref{F:Equipotential lines}). We used an electron beam starting from the origin with velocity on the $y$ axis only, which corresponds to a kinetic energy of $1~$eV. We observe very small deviations (less than $0.041\%$) that practically can be neglected, since they are smaller than the detector's resolution.} 
    \label{F:y final as function of z initial}
\end{figure}

Figure \ref{F:rho vs vy} (a) shows $y_f$ as a function of  $v_y^0$, for several different values of $U_\rho$. As expected, we observe a linear dependence of the impact position on initial velocity, and by changing $U_\rho$ (while maintaining VMI conditions, by the appropriate change to $U_z$) one can change the magnification. Figure \ref{F:rho vs vy} (b) shows the magnification as a function of $U_\rho$ which is well described by Eq. \ref{E:Magnification in VMI}.
Figure
 \ref{F:Time of Flight} (a) shows $t_f$ as function of $z_0$, in good agreement with the analytical formula (see Eq. 18 in the
supporting information). Figure \ref{F:Time of Flight} (b) shows $t_f$ as a function of $v_z^0$. In this case the difference in time-of-flight between electrons with a kinetic energy of $1~$eV in the $z$ direction to electrons produced at rest is $\simeq 30~$ns. We have also verified that $\rho_f$ has only a weak dependence on $z_0$. For example, for electrons with a kinetic energy of $1~$eV in the transverse direction, the difference in $\rho_f$  between electrons created at the center of the CSR-ReMi, and electrons created at a distance of $10~$mm away along the $\widehat{z}$-direction is less than $200~\mathrm{\mu}$m, as can be seen in Fig. \ref{F:y final as function of z initial}.

\begin{figure} 
    \centering   \includegraphics[width=.48\textwidth]{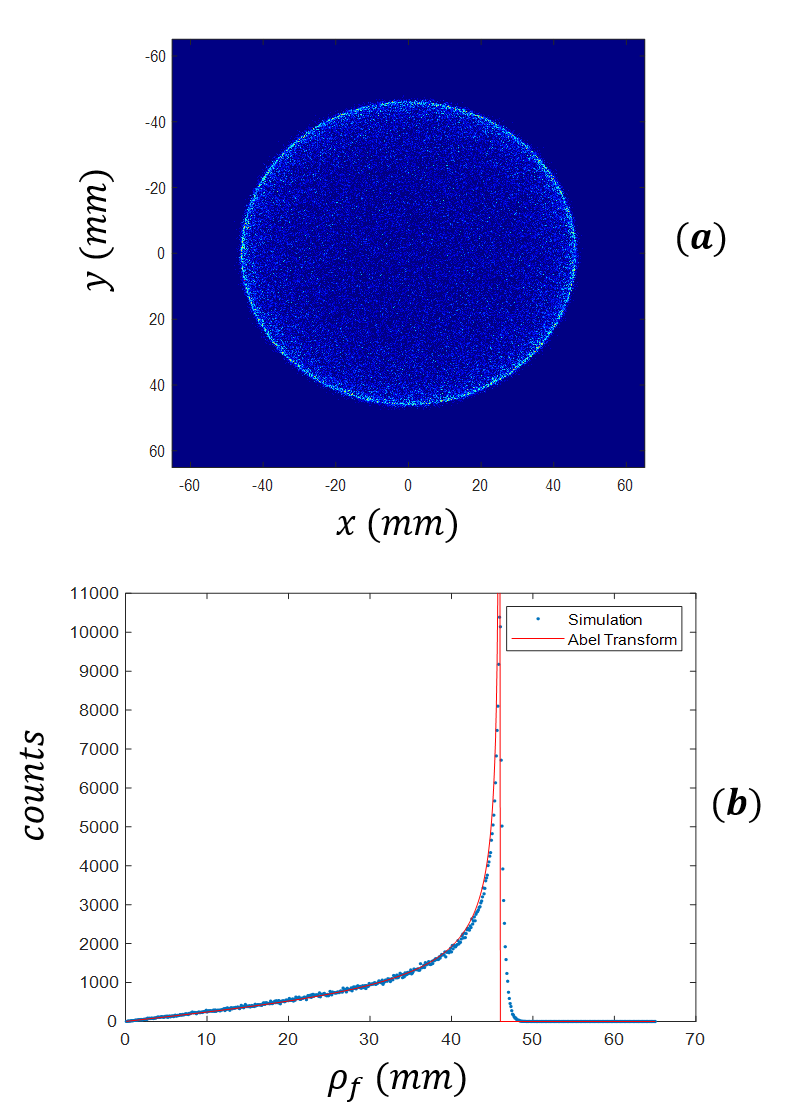}
    \caption{ \textbf{(a)} Expected image on the detector based on a SIMION simulation with 500,000 electrons with kinetic energy of $0.9~$eV, which start from a large initial volume (with a width of $80~$mm).
    \textbf{(b)} The histogram of $\rho_f$. The red line corresponds to the expected Abel transform (Eq. \ref{E:rho histogram}. In both cases values of $U_\rho=-4.26\cdot 10^{-4}~\mathrm{\frac{V}{mm^2}}$, $U_z=-0.1442~\mathrm{\frac{V}{mm}}$. were taken.}
    \label{F:Detector Image}
\end{figure}

As a demonstration of the power of the device, Fig. \ref{F:Detector Image} (a), shows the simulated image produced on the detector for the  case of $500,000$ electrons with an initial kinetic energy of $\epsilon_k=0.9~$eV, assuming the direction distribution is spherically symmetric. The electrons were emitted from a cylindrical volume with a diameter of $80~$mm (in the $\widehat{x},\widehat{y}$ directions) and were normally distributed along the $\widehat{z}$ direction with a standard deviation of $10~$mm. Figure \ref{F:Detector Image} (b) shows a histogram of $\rho_f$. The red line corresponds to the expected $\rho_f$ distribution given by:
\begin{equation} 
    I_\rho(\rho_0, \varphi)=f(\varphi)\left[ \sqrt{1-\left(\frac{\rho}{\rho_M}\right)^2}-\sqrt{1-\left(\frac{\rho+\Delta\rho}{\rho_M}\right)^2} \right]d\varphi
\end{equation} \label{E:rho histogram}
Here $\rho_M=M\sqrt{\frac{2\epsilon_k}{m}}$, and $\Delta\rho$ is the histogram's bin size. The match between this formula and the results of the simulations is good despite the formula neglecting broadening due to spherical aberrations.

\section{Position Imaging and Mixed Imaging}
\begin{figure} 
    \centering
\includegraphics[height=.6\textwidth , width=0.4\textwidth]{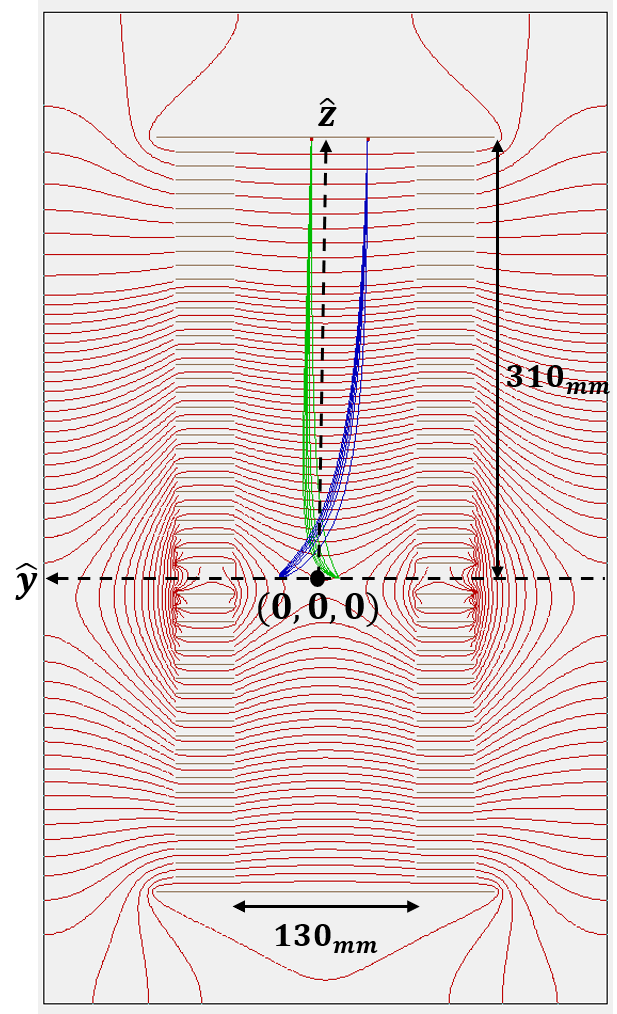}
    \caption{SIMION demonstration of position imaging in the CSR-ReMi. Here values of $U_\rho=-6.7\cdot 10^{-3}~\mathrm{\frac{V}{mm^2}}$, $Uz=-0.2~\mathrm{\frac{V}{mm}}$ are taken. The equipotential lines are shown in red. Also shown are two sets of trajectories. The blue trajectories correspond to electrons with kinetic energies in the $0-1~$eV range which all start at $y_0=-30~$mm, and the green trajectories correspond to electrons in the same kinetic energy range starting at $y_0=10~$mm. In both cases the different electrons are focused to a small spot on the detector at inverted positions ($\rho_f=30,~-10~$mm respectively).}\label{F:Position Imaging}
\end{figure}

The quadratic potential presented above (Eq. \ref{E:V}) can also be used for position imaging (also known simply as 'imaging'). In this case the goal is that $\rho_f$ depends only on $\rho_0$ and not on $v_\rho^o$. This can be achieved by setting the sine term in Eq. \ref{E:rho solution} to zero, which is the case when $\omega_\rho t_f=\pi n$ for an integer $n$. Taking $n=1$, leads to the condition:
\begin{equation}\label{E:Position Imgaing conditions}
    \frac{U_\rho}{U_z}=\frac{\beta_{pi}}{L}
\end{equation}
with:
\begin{equation}    
\beta_{pi}=\frac{1}{4} \left[ cosh(\sqrt{2}\pi)-1\right]=10.38
\end{equation}
In this case the image is inverted without magnification. Figure \ref{F:Position Imaging} shows an example of operating the CSR-ReMi in position imaging mode, where electrons produced from the same spot but with different velocities are imaged to the same spot.
\begin{figure} 
    \centering   \includegraphics[width=0.48\textwidth]{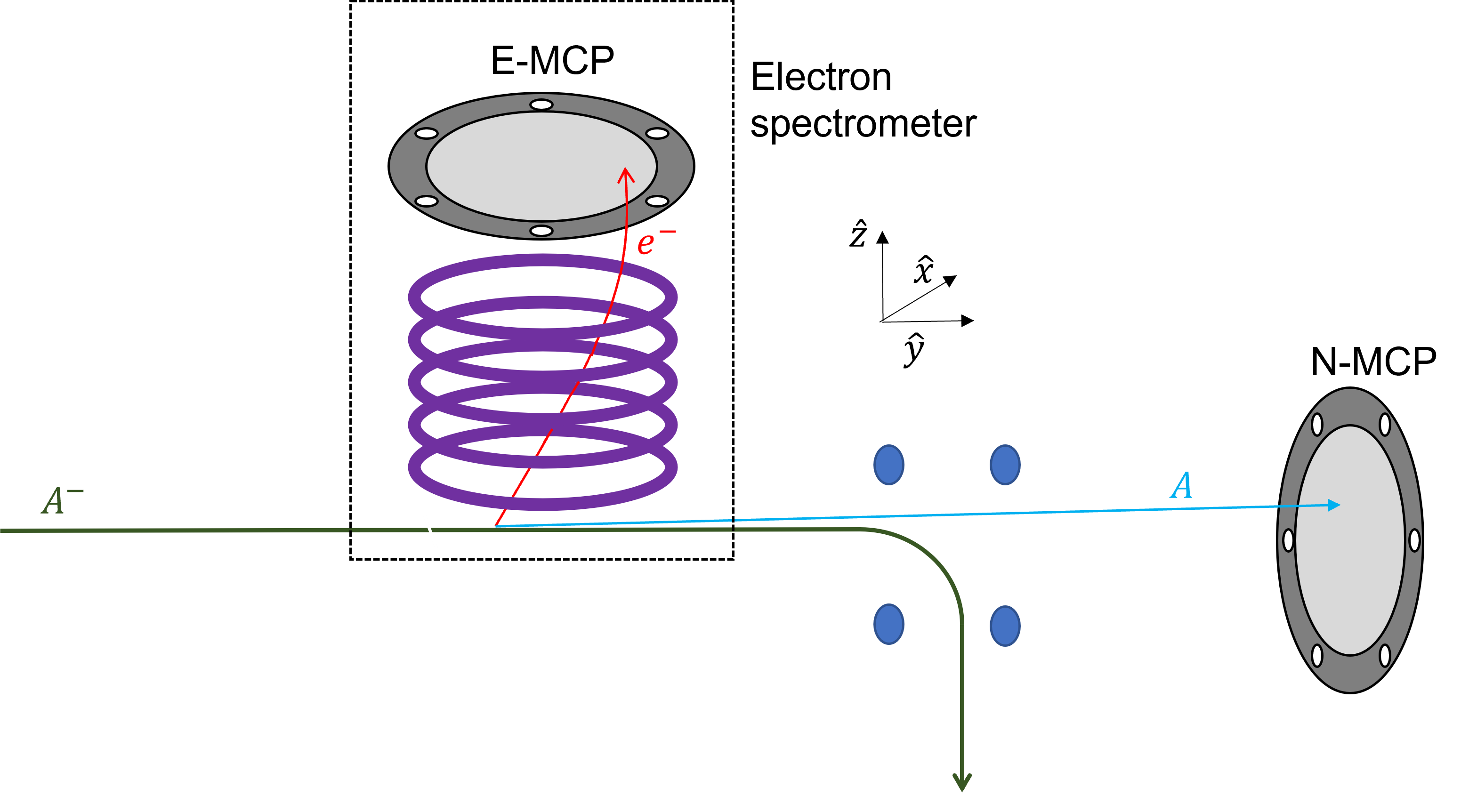}
    \caption{An illustration of a spectrometer for measuring thermionic emission from anions without prior mass selection, as an example for an application of mixed imaging: velocity map imaging along the $\widehat{x}$ direction and position imaging in the $\widehat{y}$ direction which is used to determine the position of the ionization event. This information can be used to improve the time-of-flight determination of the parent ions mass.}
    \label{F:Thermionic Spectrometer}
\end{figure}

Another interesting possibility is a mixed imaging mode with position imaging along the one axis and velocity map imaging in the perpendicular direction. In this case cylindrical symmetry is broken and therefore the implementation of this mode requires the construction of a dedicated spectrometer with elliptically shaped electrodes. Figure \ref{F:Thermionic Spectrometer} shows an illustration of a setup where such capabilities may be interesting, for studying thermionic emission without the use of mass selection. In this case a hot ion beam, $A^-$, is accelerated and directed through the electron spectrometer and then deflected into a Faraday cup. If a thermionic emission event occurs within the electron spectrometer, $A^- \rightarrow A+e^-$, then the electron will be accelerated upwards (in the $\widehat{z}$ direction) and imaged on the E-MCP detector, while the neutral fragment will continue down stream (along the $\widehat{y}$ direction) and imaged, in coincidence, on a MCP for neutrals labeled 'N-MCP' in Fig.\ref{F:Thermionic Spectrometer}. Using the time delay between the impacts on the two detectors, one can determine the mass of the parent ion, if the distance to the detector is well known. By using position imaging in the $\widehat{y}$ direction one can infer precisely the position of the ionization event and, thus infer the mass with better resolution. Using VMI in the $\widehat{x}$ position allows to determine the electrons kinetic energy distribution. Thus, one can measure the distribution of energies of thermionically emitted electrons for different ions simultaneously without the need for mass-selection. Mixed imaging can be achieved using the following solution of the Laplace equation:
\begin{equation}
V(x,y,z)=\frac{U_x}{2}x^2+\frac{U_y}{2}y^2-\frac{U_x+U_y}{2}z^2+U_z z    
\end{equation}
We define: $\omega_x=\sqrt{\frac{q U_x}{m}}$, and $\omega_y=\sqrt{\frac{q U_y}{m}}$, and require, for mixed imaging that $\omega_x t_f=\pi/2,~~\omega_y t_f=\pi$ and  $\omega_y=2\omega_x$ and therefore that:
\begin{equation}
    U_y=4U_x
\end{equation}
In this case $x_f=-Mv_x^0$ where the magnification is $M=\frac{1}{\omega_x}$ and $y_f=-y_0$. For $z_0=v_z^0=0$ the solution for the equation of motion in the $\widehat{z}$ direction is given by:
\begin{equation}
z(t)=\frac{U_z}{U_x+U_y}\left[1-cosh\left(\sqrt{q\frac{U_x+U_y}{m}}t\right)\right]    
\end{equation}
using $t_f=\frac{\pi}{2\omega_x}$ leads to an equation for $U_z$:
\begin{equation}
U_z=\frac{5LU_x}{1-cosh(\pi \sqrt{\frac{5}{4}})} 
\end{equation}
Thus, given $\epsilon_{max}$ (the maximal kinetic energy to be measured) $U_x,~U_y$ and $U_z$ should be set according to:
\begin{equation}
    U_x=\frac{2\epsilon_{max}}{qR_d^2}=\frac{U_y}{4}=-3.16 \frac{U_z}{L}
\end{equation}.

\begin{figure} 
    \centering
\includegraphics[width=.3\textwidth]{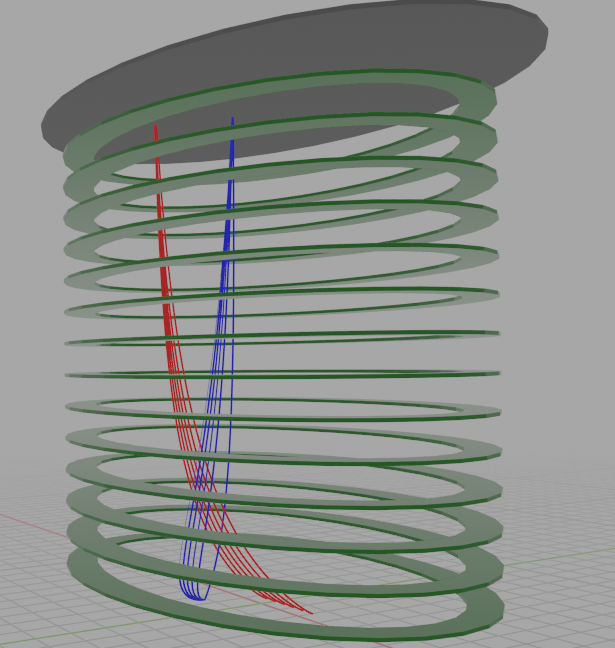}
    \caption{An illustration of mixed imaging. We chose here $U_x=-3.75\cdot 10^{-3}~\mathrm{\frac{V}{mm^2}}$, $U_y=-0.015~\mathrm{\frac{V}{mm^2}}$ and $U_z=0.1782~\mathrm{\frac{V}{mm}}$. The red trajectories correspond to electrons starting from the same $y_0$ and having the same $v_x^0$ (corresponding to a kinetic energy of $1~$eV) but with a range of different $x_0$. The blue trajectories describe electrons starting at $y_0=6~$mm with different $v_y^0$ which correspond to kinetic energies in the $0-1.25~$eV range. Both sets of trajectories are focused to a small spot on the detector.}\label{F:Mixed Imaging trajectories}
\end{figure}

Figure \ref{F:Mixed Imaging trajectories} shows an example of the results of a SIMION simulation of this kind of a device. Here a stack of elliptical electrodes with $R_x=40~$mm and $R_y=20~$mm is used. The red lines correspond to the trajectories of a series of electrons starting at different initial positions along the $\widehat{x}$ axis and focused to one point, while the blue trajectories are a series of electrons starting from the same point but having different $v_y^0$  velocities, which are also focused to a single spot.

\section{Conclusion}
In conclusion, we have presented an analytical derivation showing how a quadratic potential can be used for velocity map or position imaging with no spherical aberrations. We have used SIMION simulations to demonstrate the applicability of these derivations to a practical device: as a means of operating the CSR-ReMi in a VMI mode so that it can be used for photo-electron spectroscopy. The simulations suggest $40$-fold improvement of the focusing factor over the present state of the art. We have also shown how this methodology can be used in a mixed imaging mode and discussed one possible use for such a device.

\section{Acknowledgments}
We are grateful to Oded Heber for insightful conversations and for suggesting to use the CSR-ReMi in VMI mode. This work  was supported by the Israeli Science Foundation, the Minerva center for making bonds by fragmentation and by COST Action CA18212 (Molecular Dynamics in the GAS phase (MD-GAS)).

\bibliographystyle{naturemag}
\bibliography{MainBib}
\end{document}